\documentclass[pra,showpacs,amsmath,amssymb,twocolumn, 10pt]{revtex4}

\usepackage{amsthm}
\usepackage{dcolumn}
\usepackage{bm}
\usepackage{graphicx}

\begin{document}

\newtheorem{corollary}{Corollary}
\newtheorem{definition}{Definition}
\newtheorem{example}{Example}
\newtheorem{lemma}{Lemma}
\newtheorem{proposition}{Proposition}
\newtheorem{theorem}{Theorem}
\newtheorem{fact}{Fact}
\newtheorem{property}{Property}
\newcommand{\bra}[1]{\langle #1|}
\newcommand{\ket}[1]{|#1\rangle}
\newcommand{\braket}[3]{\langle #1|#2|#3\rangle}
\newcommand{\ip}[2]{\langle #1|#2\rangle}
\newcommand{\op}[2]{|#1\rangle \langle #2|}

\newcommand{\tr}{{\rm tr}}
\newcommand {\E } {{\mathcal{E}}}
\newcommand {\F } {{\mathcal{F}}}
\newcommand {\diag } {{\rm diag}}
\newcommand{\slocc}{\overset{\underset{\mathrm{SLOCC}}{}}{\longrightarrow}}

\title{\Large {\bf The Tensor Rank of the Tripartite State $\ket{W}^{\otimes n}$}}
\author{Nengkun Yu$^1$}
\email{nengkunyu@gmail.com}
\author{Eric Chitambar$^2$}
\email{echitamb@alumni.nd.edu}
\author{Cheng Guo}
\email{cheng323232@gmail.com}
\author{Runyao Duan$^1$}
\email{dry@tsinghua.edu.cn}
\affiliation{$^1$Department of Computer Science and Technology, Tsinghua
University, Beijing 100084, China and\\ Center for Quantum Computation
and Intelligent Systems (QCIS), Faculty of Engineering and
Information Technology, University of Technology, Sydney, NSW 2007,
Australia
\\
$^2$Physics Department, University of Michigan, 450 Church Street,
Ann Arbor, Michigan 48109-1040, USA.}

\begin{abstract}
Tensor rank refers to the number of product states needed to express
a given multipartite quantum state. Its non-additivity as an
entanglement measure has recently been observed. In this brief
report, we estimate the tensor rank of multiple copies of the
tripartite state
$\ket{W}=\tfrac{1}{\sqrt{3}}(\ket{100}+\ket{010}+\ket{001})$. Both an
upper bound and a lower bound of this rank are derived. In
particular, it is proven that the rank of $\ket{W}^{\otimes 2}$ is
seven, thus resolving a previously open problem. Some implications
of this result are discussed in terms of transformation rates
between $\ket{W}^{\otimes n}$ and multiple copies of the state
$\ket{GHZ}=\tfrac{1}{\sqrt{2}}(\ket{000}+\ket{111})$.
\end{abstract}

\pacs{03.65.Ud, 03.67.Hk}

\maketitle

With quantum entanglement being a proven asset to information
processing and computational tasks, much effort has been devoted to
quantifying it as a resource.  One particular entanglement measure
for a general pure state $\ket{\psi}\in H_1\otimes H_2\otimes...
H_n$ is the tensor rank which, denoted by $rk(\ket{\psi})$, is
defined as the minimum number $r$ such that there exist states
$\ket{\phi_{ik}}\in H_k$ ($1\leq i\leq r$ and $1\leq k\leq n$) for
which
\begin{equation*}
\ket{\psi}=\sum_{k=1}^r
\ket{\phi_{i1}}\ket{\phi_{i2}}\cdot\cdot\cdot\ket{\phi_{in}}.
\end{equation*}
Tensor rank is a legitimate entanglement measure since it is a
strictly non-increasing quantity under local operations (even
probabilistically).  Another key feature is that it provides a
method of detecting entanglement: a multipartite pure state is
entangled if and only if its tensor rank is larger than one.
Further properties of the tensor rank and its application to quantum
entanglement have been studied by Eisert and co-workers in Ref.
\cite{Eisert-2001a} and \cite{Hein-2004a}. For bipartite systems,
tensor rank is equivalent to the so-called ``Schmidt rank''.

One way of partitioning a multipartite state space is as follows:
two states are considered to be equivalent if and only if they can
be reversibly converted from one to the other by operations
belonging to the class of Stochastic Local Operations with Classical
Communication (SLOCC). The ability to transform a state $\ket{\psi}$
to a state $\ket{\phi}$ with SLOCC is symbolically expressed as
$\ket{\psi}\slocc\ket{\phi}$. It is well-known that in bipartite
systems, $\ket{\psi}\slocc\ket{\phi}$ if and only if
$rk(\ket{\psi})\geq rk(\ket{\phi})$.  Consequently, SLOCC
convertibility induces a total ordering among bipartite pure states
that can be completely characterized by the Schmidt rank.  Another
nice property of the Schmidt rank is that it is additive in the
sense that $\log_2(rk(\ket{\psi}^{\otimes n}))=n\cdot
\log_2(rk(\ket{\psi}))$. In contrast, the general multipartite tensor
rank is insufficient for determining SLOCC equivalence
\cite{Verstraete-2002a}, and it is not an additive quantity
\cite{Chitambar-2008a}.

The main purpose of this brief report is to evaluate the tensor rank of
the tripartite state $\ket{W}^{\otimes n}$. As the rank is unchanged
by overall scalar multiplications, from now on we will work with
unnormalized states. D$\mathrm{\ddot{u}}$r \textit{et al.\!} \cite{Dur-2000a} were the
first to observe that within three qubit systems, there exist two
distinct equivalence classes of genuinely tripartite entangled
states, one represented by the state $\ket{GHZ}$ and the other
represented by $\ket{W}$. While single copy transformations are not
possible between $\ket{W}$ and $\ket{GHZ}$, a natural question is:
how many copies of $\ket{W}$ are needed to obtain a single
$\ket{GHZ}$ by SLOCC, and vice versa?  More precisely, what are the
lowest possible ratios $\tfrac{m}{n}$ for the respective
transformations $\ket{GHZ}^{\otimes m}\slocc\ket{W}^{\otimes n}$ and
$\ket{W}^{\otimes m}\slocc\ket{GHZ}^{\otimes n}$?  As noted above,
the tensor rank is monotonically decreasing under SLOCC. Hence a way
of evaluating these ratios is to calculate the tensor rank for the
above multiple-copy states.

In Ref. \cite{Chitambar-2008a}, an eight-term product state
expansion for $\ket{W}^{\otimes 2}$ was found. This indicates that
$rk(\ket{W}^{\otimes 2})\leq 8$, and the transformation
$\ket{GHZ}^{\otimes 3}\slocc\ket{W}^{\otimes 2}$ is feasible. In
this brief report, we determine the exact value $rk(\ket{W}^{\otimes
2})=7$. As an immediate consequence, we have the improved upper
bound of $rk(\ket{W}^{\otimes n})\leq 7^{\lceil n/2\rceil}$. On the
other hand, we find a lower bound of $rk(\ket{W}^{\otimes n})\geq
2^{n+1}-1$. It is obvious that this lower bound is tight when $n=2$.

In the following, we will use $\mathrm{Tr}_{i}(\cdot)$ to stand for
a partial trace with respect to subsystem $i$, and the support of a
positive operator will refer to its range. We will need the
following lemma from Ref. \cite{Chitambar-2008a}, which gives an
alternative description of tensor rank.
\begin{lemma}\label{lemma1}\upshape
Suppose $\ket{\psi}\in H_1\otimes \cdot\cdot\cdot\otimes H_n$. For
an arbitrarily chosen $i$, $rk(\ket{\psi})$ equals the minimum
number of product states in  $H_1\otimes\cdot\cdot\cdot\otimes
H_{i-1}\otimes H_{i+1}\otimes\cdot\cdot\cdot\otimes H_n$ whose
linear span contains the support of
$\mathrm{Tr}_{i}(\op{\psi}{\psi})$.
\end{lemma}

For the state $\ket{W}^{\otimes 2}\in H_A\otimes H_B\otimes H_C$, let $\rho=\mathrm{Tr}_{A}(\op{{W}^{\otimes 2}}{{W}^{\otimes 2}})$,  the support of $\rho$ is spanned by the states
$$\{\ket{00},\ket{01}+\ket{10},\ket{02}+\ket{20},\ket{03}+\ket{12}+\ket{21}+\ket{30}\},$$
for $0\leq i,j<4$, $\ket{ij}=\ket{i}_B\ket{j}_C$ , $\ket{i}_B$ and $\ket{j}_C$ denote the two-qubits' state of B-part and C-part, respectively.(For simplicity, here we assume that
$\ket{0}=\ket{00}$, $\ket{1}=\ket{01}$, $\ket{2}=\ket{10}$, and
$\ket{3}=\ket{11}$).

One can easily check that this set is contained in the linear span
of $$\{\ket{00},
\ket{03},\ket{30},\ket{\Psi_1}\ket{\Psi_1},\ket{\Psi_2}\ket{\Psi_2},\ket{\Psi_3}\ket{\Psi_4},
\ket{\Psi_4}\ket{\Psi_4}\}$$
 where $\ket{\Psi_1}=\ket{0}+\ket{1}+\ket{2}$, $\ket{\Psi_2}=\ket{0}+\ket{1}-\ket{2}$, $\ket{\Psi_3}=\ket{1}+\ket{2}$,
 and $\ket{\Psi_4}=\ket{1}-\ket{2}$.  Indeed, we have:
\begin{eqnarray*}
\ket{00}+\ket{01}+\ket{10}&=&(\ket{\Psi_1}\ket{\Psi_1}-\ket{\Psi_3}\ket{\Psi_3}\\
&&+\ket{\Psi_2}\ket{\Psi_2}-\ket{\Psi_4}\ket{\Psi_4})/2,\\
\ket{02}+\ket{20}&=&(\ket{\Psi_1}\ket{\Psi_1}-\ket{\Psi_3}\ket{\Psi_3}\\
&&-\ket{\Psi_2}\ket{\Psi_2}+\ket{\Psi_4}\ket{\Psi_4})/2,\\
\ket{12}+\ket{21}&=&(\ket{\Psi_3}\ket{\Psi_3}-
\ket{\Psi_4}\ket{\Psi_4})/2.
\end{eqnarray*}
Lemma \ref{lemma1} implies that $rk(\ket{W}^{\otimes 2})\leq 7$.
Using the above equalities one obtains a seven term decomposition:
\begin{eqnarray*}
\ket{W}^{\otimes 2}&=&\ket{300}+\ket{030}+\ket{00}(\ket{3}-\ket{2})\\
&+&\ket{\Psi_1}\ket{\Psi_1}(\ket{0}-\ket{1}-\ket{2})/2\\
&+&\ket{\Psi_2}\ket{\Psi_2}(\ket{1}-\ket{0}-\ket{2})/2\\
&+&\ket{\Psi_3}\ket{\Psi_3}(\ket{1}+\ket{2})/2+\ket{\Psi_4}\ket{\Psi_4}(\ket{2}-\ket{1})/2.
\end{eqnarray*}

We now proceed to prove a lower bound for the tensor rank of $n$
copies of $\ket{W}$.  By Lemma \ref{lemma1}, it is sufficient to
evaluate the minimum number of product states whose linear span
contains the set $$\{\ket{00},\ket{01}+\ket{10}\}^{\otimes
n}=\{\ket{\varphi_i}|0\leq i\leq 2^n-1\}.$$ Without loss of generality, assume that
$\ket{\varphi_0}=\ket{0}^{\otimes n}$ and
$|\varphi_{2^n-1}\rangle=(|01\rangle+|10\rangle)^{\otimes n}$.
 The essential piece in the
following lemma is the fact that the state
$$\ket{\varphi_{2^n-1}}+\sum_{i=0}^{2^n-2} \alpha_i\ket{\varphi_i}$$
always has a Schmidt rank of $2^n$ for any $\alpha_i\in\mathbb{C}$.
This can be easily verified by calculating the matrix rank after
taking a partial trace.

\begin{lemma}\label{lemma2}\upshape
 $rk(\ket{W}^{\otimes n})\geq 2^{n+1}-1$.
\end{lemma}

\textbf{Proof:} Assume that there exist $2^{n+1}-2$ product states
$\{\ket{\xi_i}|0\leq i\leq 2^{n+1}-3\}$ whose linear span contains
$\{\ket{\varphi_i}|0\leq i\leq 2^n-1\}$. Without loss of generality,
let  $\ket{\xi_0}=\ket{\varphi_0}$.  For $1\leq i\leq 2^n-1$, put
$$|\varphi_i\rangle=\sum_{k=0}^{2^{n+1}-3}\beta_{ik}|\xi_k\rangle.$$
Since the matrix $B=(\beta_{ik})$ for $1\leq i\leq 2^n-2$ and $0\leq
k\leq 2^{n+1}-3$ has rank $2^n-2$, there exist $2^n-2$ linear
independent columns ${l_1,l_2\cdot\cdot\cdot l_{2^n-2}}$. Then, one
can find $2^n-2$ complex number ${\delta_1,\delta_2\cdot\cdot\cdot
\delta_{2^n-2}}$ such that $\mu_{l_i}=0$ for all $1\leq i\leq
{2^n-2}$ and $$\ket{\varphi_{2^n-1}}+\sum_{i=1}^{2^n-2} \delta_i
\ket{\varphi_i}= \sum_{i=0}^{2^{n+1}-3}\mu_{l _i}\ket{\xi_i}.$$
Consequently, we see that $$\ket{\varphi_{2^n-1}}+\sum_{i=1}^{2^n-2}
\delta_i \ket{\varphi_i}+ \mu_{0}\ket{\varphi_0}=
\sum_{i=1}^{2^{n+1}-3}\mu_{i}\ket{\xi_i}$$ is a
$2^{n+1}-3-(2^n-2)=2^n-1$ product state expansion.  This, however,
is impossible since $\ket{\varphi_{2^n-1}}$ has a Schmidt rank of
$2^n$. \hfill $\blacksquare$

In light of our seven term construction for $\ket{W}^{\otimes 2}$,
we now have the following
\begin{lemma}\label{lemma3}\upshape
$rk(|W^{\otimes 2}\rangle)=7$.
\end{lemma}

Combining the above lemmas as well as some of their immediate
consequences, we obtain the main results of this paper and state
them in the following theorem.  Part (b) relies on the facts that $rk(\ket{GHZ}^{\otimes n})=2^n$ and that a non-increase in tensor rank is sufficient for convertibility from $\ket{GHZ}^{\otimes n}$.
\begin{theorem}
\begin{eqnarray*}
&(a)&\;\;2^{n+1}-1\leq rk(\ket{W}^{\otimes n})\leq\begin{cases} 7^{n/2}\;\;\text{\upshape even}\;n\\3\cdot 7^{\tfrac{n-1}{2}}\;\;\text{\upshape odd}\;n,\end{cases}\\
&(b)&\;\; \frac{m}{n+1}\geq \tfrac{1}{2}\log_2 7 \Rightarrow \ket{GHZ}^{\otimes m}\slocc\ket{W}^{\otimes n},\\
&(c)&\;\; \ket{W}^{\otimes m}\slocc\ket{GHZ}^{\otimes n}\Rightarrow \frac{n}{m+1}\leq \tfrac{1}{2}\log_2 7.
\end{eqnarray*}
\end{theorem}

Let $N=2^n$, part (a) of the above theorem can be restated as
follows: $2N-1\leq rk(\ket{W}^{\otimes log_2N})\leq O(N^\nu)$ where
$\nu=\tfrac{\log_27}{2}$.   There exists an interesting similarity
between this and the bounds on the tensor rank of the state
$\ket{\Phi^3}^{\otimes \log_2N}$ where
$\ket{\Phi^3}=\ket{\Phi}_{AB}\ket{\Phi}_{AC}\ket{\Phi}_{BC}$ and
$\ket{\Phi}_{ij}=\ket{0_i0_j}+\ket{1_i1_j}$.  In Ref.
\cite{Chitambar-2008a} it was shown that $\tfrac{5}{2}N^2-3N\leq
rk(\ket{\Phi^3})^{\otimes \log_2N}\leq O(N^\omega)$ where
$\omega=2.36$.  It turns out that $\omega$ corresponds to the
so-called ``exponent for matrix multiplication'' which is the
smallest real number such that an algorithm exists for multiplying
two $N\times N$ matrices using $O(N^\omega)$ multiplications.
Extensive work has been devoted to determining the exact value of
$\omega$ and most researchers hypothesize its value to be two.  In a
similar manner, we can define the ``exponent for the W-state'' as
the smallest real number $\nu$ such that $rk(\ket{W}^{\otimes
\log_2N})\leq O(N^\nu)$.  We conjecture that $\nu=1$, and if this
were true, then for any $\epsilon >0$, there would exist some $n$
such that $\ket{GHZ}^{\otimes \lfloor
n(1+\epsilon)\rfloor}\slocc\ket{W}^{\otimes n}$.  It remains an open
challenge to verify whether or not this speculation is correct.

We are grateful to Lin Chen for pointing out one mistake in Theorem 1 of our pervious version.
We thank Prof. Mingsheng Ying for useful comments.
This work was partly supported by the Natural Science Foundation of
China (Grant Nos.  60736011, 60702080, and 60621062), the Hi-Tech
Research and Development Program of China (863 project) (Grant No.
2006AA01Z102), and the National Basic Research Program of China
(Grant No. 2007CB807901).  Additional support for E.C. came from the National
Science Foundation of the United States under
Awards 0347078 and 0622033.


\begin{thebibliography}{99}

\bibitem{Eisert-2001a} J. Eisert and H. J. Briegel, Phys. Rev. A \textbf{64}, 022306
(2001).

\bibitem{Hein-2004a} M. Hein, J. Eisert, and H. J. Briegel, Phys. Rev. A \textbf{69},
062311 (2004).

\bibitem{Verstraete-2002a} F. Verstraete, J. Dehaene, B. De Moor, and H. Verschelde,
Phys. Rev. A \textbf{65}, 052112 (2002).

\bibitem{Chitambar-2008a} E. Chitambar, R. Duan, and Y. Shi, Phys. Rev. Lett. \textbf{101},
140502 (2008).

\bibitem{Dur-2000a} W. D$\mathrm{\ddot{u}}$r, G. Vidal, and J. I. Cirac, Phys. Rev. A \textbf{62}, 062314
(2000).

\end{thebibliography}
\end{document}